\documentclass[epj]{webofc}
\usepackage[varg]{txfonts}   % Web of Conferences font
\usepackage{amsmath,amssymb,bm}
\renewcommand\d{\partial}
\wocname{EPJ Web of Conferences}
\woctitle{CONF12}
\begin{document}
\selectlanguage{english}
\title{Chiral transport of neutrinos in supernovae}

\author{Naoki Yamamoto\inst{1}
\fnsep\thanks{\email{nyama@rk.phys.keio.ac.jp}}}

\institute{Department of Physics, Keio University, Yokohama 223-8522, Japan}

\abstract{The conventional neutrino transport theory for core-collapse supernovae misses one key property of neutrinos: the left-handedness. The chirality of neutrinos modifies the hydrodynamic behavior at the macroscopic scale and leads to topological transport phenomena. We argue that such transport phenomena should play important roles in the evolution of core-collapse supernovae, and, in particular, lead to a tendency toward the inverse energy cascade from small to larger scales, which may be relevant to the origin of the supernova explosion.}
\maketitle
\section{Introduction}
\label{intro}
The core-collapse supernova explosion, which is the transition of a massive star into a neutron star, is one of the most energetic explosions in the Universe. Phenomenologically, this explosion is important to release various elements produced by nuclear fusions inside the massive star into the space. Theoretically, however, the physical mechanism of core-collapse supernova explosion has not been fully understood so far. When a massive star collapses, most of the gravitational binding energy released is carried away by neutrinos. In order to figure out the origin of the supernova explosion, it is thus essential to treat the neutrino transport appropriately. Currently, numerical simulations of the conventional neutrino transport theory in three dimensions (3D) have difficulty in reproducing a supernova explosion with the observed explosion energy; see, e.g., Ref.~\cite{Janka:2012wk}. For this reason, unraveling the mechanism of the supernova explosion has still been one of the long-standing problems in astrophysics. 

Yet one key property of neutrinos has been missed in the conventional neutrino transport theory for supernovae: the left-handedness of neutrinos. It has been put forward by the present author that the {\it microscopic} parity violation due to the chirality of neutrinos must be reflected in the {\it macroscopic} evolution of supernovae \cite{Yamamoto:2015gzz, Yamamoto:2016xtu}. Since parity is violated in the microscopic theory, parity must also be violated in the kinetic theory (and hydrodynamics if applicable) for left-handed neutrinos. Although such chiral effects have been completely ignored in the conventional transport theory in astrophysics, the modified kinetic theory and hydrodynamics that take into account these effects have been constructed in the areas of nuclear and particle physics and are now called the chiral kinetic theory \cite{Son:2012wh, Stephanov:2012ki, Son:2012zy, Chen:2012ca} and chiral (or anomalous) hydrodynamics \cite{Son:2009tf}, respectively. 

In fact, the supernova is the system with the {\it largest global parity violation} in the Universe. There a huge amount of left-handed neutrinos are produced in the process of electron capture: ${\rm p} + {\rm e}_{\rm L} \rightarrow {\rm n} + \nu_{\rm L}$. One then expects that dynamical evolution of this parity-violating system should be qualitatively different from the evolution of usual parity-invariant systems (like the air and water). 

In the following, we will ignore the neutrino mass for simplicity, as it is much smaller than its typical energy scale in supernovae.

\section{Chiral transport phenomena and chiral kinetic theory}
\label{sec:kinetics}

\subsection{Chiral vortical effect}
\label{sec:CVE}
One typical example of the modification to the usual transport phenomena because of the chirality of neutrinos is the so-called chiral vortical effect  (CVE). This was first discovered by Vilenkin in the context of neutrino emission from a rotating black hole \cite{Vilenkin:1979ui}, and it has been recently rediscovered by the application of the gauge-gravity duality \cite{Erdmenger:2008rm, Banerjee:2008th}. 
The CVE is the contribution to the particle number current in the direction of the vorticity ${\bm \omega} = {\bm \nabla} \times {\bm v}$, where ${\bm v}$ is the fluid velocity. For the left-handed neutrino matter at finite chemical potential $\mu$ and temperature $T$, it is given by
\begin{equation}
\label{CVE}
{\bm j} = -\left(\frac{\mu^2}{8\pi^2} + \frac{T^2}{24} \right) {\bm \omega}\,.
\end{equation}
Physically, the CVE can be understood as follows \cite{Kharzeev:2015znc} (although this argument is just schematic). In the presence of vorticity, the spin of the fermion tends to preferably align in the direction of the vorticity due to the spin-orbit coupling. By the definition of chirality, momentum of left-handed fermion aligns in the opposite direction of the vorticity, meaning that left-handed neutrinos tend to move in the opposite direction of the vorticity. Therefore, the particle number current flows in the opposite direction of the vorticity in neutrino matter. This is the CVE. 

The precise transport coefficient of the CVE above can be computed using the microscopic quantum field theory \cite{Vilenkin:1979ui, Landsteiner:2011cp} or chiral kinetic theory \cite{Stephanov:2012ki, Chen:2014cla, Chen:2015gta}. Apparently, the CVE is a consequence of the chirality of neutrinos, and it is absent, e.g., in the air. This example clearly shows that the transport properties of neutrino matter should be drastically different from those of usual parity-invariant matter.

\subsection{Berry curvature}
Notice that the conventional Boltzmann equation cannot describe the CVE, simply because it cannot distinguish between right- and left-handedness. This means that the conventional Boltzmann equation has some deficiency for left-handed neutrinos. We thus need to modify the framework of kinetic theory to incorporate such chiral effects. As found in Ref.~\cite{Son:2012wh}, the essential ingredient is the so-called Berry curvature, which is related to some topological nature of a system.

That the chirality of fermions is related to topology can be understood as follows. Consider a Fermi surface of right-handed fermions at zero temperature. Recall that the direction of momentum is always the same as the direction of the spin. So if the end point of the momentum vector ${\bm p}$ covers the whole Fermi surface, then the end point of the spin vector ${\bm s}$ also covers a sphere in spin space. Namely, there is a mapping from the Fermi sphere $S^2$ in momentum space to the sphere $S^2$ in spin space. This suggests the presence of a nontrivial topology characterized by the winding number $+1$ for right-handed fermions. For left-handed fermions, on the other hand, the direction of momentum is always opposite as the direction of the spin, and the winding number is $-1$. Hence, chirality can be regarded as a topological invariant.

The effects of this topology can be captured by the Berry curvature in the dynamics. The winding number above corresponds to the presence of the singularity of a monopole at the origin in momentum space with the charge $\pm 1$ for right- and left-handed chiral fermions, respectively. The Berry curvature is exactly the ``magnetic field" of this monopole in momentum space, expressed by 
\begin{equation}
\label{Omega}
{\bm \Omega}_{\bm p} = \pm \frac{\bm p}{2|{\bm p}|^3}\,.
\end{equation}
So the neutrino matter has a nontrivial Berry curvature and it is a {\it topological matter}; its nonequilibrium dynamics is affected by the topology that originates from chirality of neutrinos, as we will explicitly show below. 

To keep our argument generic, we first consider a {\it charged} chiral fermion. In the electromagnetic fields, the action is given by \cite{Son:2012wh, Stephanov:2012ki, Son:2012zy}
\begin{equation}
\label{S}
S = \int \left[({\bm p} + {\bm A}) \cdot {\rm d}{\bm x} - (\epsilon_{\bm p} + \phi) {\rm d}t - {\bm a}_{\bm p} \cdot {\rm d}{\bm p} \right]\,.
\end{equation}
Here ${\bm A}$ and $\phi$ are the vector and scalar potentials in coordinate space, ${\bm a}_{\bm p}$ is the Berry connection (or the ``gauge field" in momentum space) which is related to the Berry curvature by ${\bm \Omega}_{\bm p} = {\bm \nabla}_{\bm p} \times {\bm a}_{\bm p}$, and
\begin{equation}
\epsilon_{\bm p} = |{\bm p}|(1 - {\bm \Omega}_{\bm p} \cdot {\bm B})
\end{equation}
is the dispersion relation of the chiral fermion \cite{Son:2012zy, Chen:2014cla}. The Berry curvature correction of the dispersion relation here accounts for the effect of the magnetic moment of the chiral fermion. From the action (\ref{S}), one can write down the equations of motion for the chiral fermion as
\begin{align}
\dot {\bm x} &= \tilde {\bm v} + \dot {\bm p} \times {\bm \Omega}_{\bm p}\,,
\\
\dot {\bm p} &= \tilde {\bm E} + \dot {\bm x} \times {\bm B}\,,
\end{align}
where $\tilde {\bm v} \equiv {\d \epsilon_{\bm p}}/{\d {\bm p}}$ and $\tilde {\bm E} \equiv {\bm E} - {\d \epsilon_{\bm p}}/{\d {\bm x}}$. Solving these simultaneous equations in terms of $\dot {\bm x}$ and $\dot {\bm p}$ lead to
\begin{align}
\label{xdot}
\dot {\bm x} &= (1 + {\bm B} \cdot {\bm \Omega}_{\bm p})^{-1} \left[\tilde {\bm v} + \tilde {\bm E}  \times {\bm \Omega}_{\bm p} + (\tilde {\bm v} \cdot {\bm \Omega}_{\bm p}){\bm B} \right]\,,
\\
\label{pdot}
\dot {\bm p} &= (1 + {\bm B} \cdot {\bm \Omega}_{\bm p})^{-1}  \left[\tilde {\bm E} + \tilde {\bm v} \times {\bm B} + (\tilde {\bm E} \cdot {\bm B}) {\bm \Omega}_{\bm p} \right]\,.
\end{align}

\subsection{Chiral kinetic theory}
Once the modified equations of motion are understood, one can easily obtain the chiral kinetic theory: substituting Eqs.~(\ref{xdot}) and (\ref{pdot}) into the Boltzmann equation in terms of the distribution function $f(t, {\bm x}, {\bm p})$,
\begin{equation}
\frac{{\rm d}f}{{\rm d}t} = \frac{\d f}{\d t} + \dot {\bm x} \cdot \frac{\d f}{\d {\bm x}} + \dot {\bm p} \cdot \frac{\d f}{\d {\bm p}} = C[f]\,,
\end{equation}
where $C[f]$ is the collision term, we arrive at \cite{Son:2012wh, Stephanov:2012ki, Son:2012zy}
\begin{equation}
 \frac{\d f}{\d t} +  (1 + {\bm B} \cdot {\bm \Omega}_{\bm p})^{-1} \left(
 \left[\tilde {\bm v} + \tilde {\bm E}  \times {\bm \Omega}_{\bm p} + (\tilde {\bm v} \cdot {\bm \Omega}_{\bm p}){\bm B} \right] \cdot \frac{\d f}{\d {\bm x}} 
 + \left[\tilde {\bm E} + \tilde {\bm v} \times {\bm B} + (\tilde {\bm E} \cdot {\bm B}) {\bm \Omega}_{\bm p} \right] \cdot \frac{\d f}{\d {\bm p}} \right)
 = C[f]\,.
\end{equation}
This is the chiral kinetic theory for charged chiral fermions. This kinetic theory can distinguish between right- and left-handed fermions as the signs of the Berry curvature in Eq.~(\ref{Omega}) are opposite between the two. 

As the consequences of the Berry curvature corrections above, the expressions of the current and stress energy-momentum tensor are also modified. From now on, let us focus on the charge neutral neutrinos, with keeping our application to supernovae in mind. Then the current and energy-momentum tensor are given by \cite{Son:2012zy, Chen:2015gta}
\begin{align}
\label{j_kinetics}
j^i &= \int \frac{{\rm d}^3 {\bm p}}{(2\pi)^3} \left(\hat p^i f - p \epsilon^{ijk}\Omega_{\bm p}^j \d_k f \right)\,,
\\
\label{T_kinetics}
T^{ij} &= \int \frac{{\rm d}^3 {\bm p}}{(2\pi)^3} \left(\hat p^i \hat p^j f - \frac{1}{2}p^i \epsilon^{jkl} \Omega_{\bm p}^k \d_l f 
- \frac{1}{2}p^j \epsilon^{ikl} \Omega_{\bm p}^k \d_l f \right)\,,
\end{align}
where $p \equiv |{\bm p}|$ and $\hat {\bm p} \equiv {\bm p}/p$. (Other components of $j^{\mu}$ and $T^{\mu \nu}$ are the same as usual ones.)

In the realistic situation of supernovae, there are not only neutrinos, but also ordinary matter like nuclei and electrons, which must also be taken into account in our transport theory. In this case, the full transport equations are given by the total energy-momentum conservations and total current conservation (together with the anomalous effects). 
% $\d_{\alpha} (T_{\rm hyd}^{\alpha \beta} + T_{\nu}^{\alpha \beta}) = 0$. Here $T_{\rm hyd}^{\alpha \beta}$ is the energy-momentum tensor of fluids for ordinary matter (nuclei and electrons), and $T_{\nu}^{\alpha \beta}$ is the energy-momentum tensor of neutrinos. 
Note here that one can use the hydrodynamics for ordinary matter, which is thermalized at the astrophysical length scale. On the other hand, one cannot always use the hydrodynamics for neutrinos outside the core of supernovae where neutrinos are not thermalized. For neutrinos, one thus needs to use the current (\ref{j_kinetics}) and energy-momentum tensor (\ref{T_kinetics}) in the chiral kinetic theory. This is the idea of the neutrino radiation {\it chiral} hydrodynamics \cite{radiation}.

\section{Application to core-collapse supernovae}

\subsection{Neutrino matter as a chiral quantum liquid}
In the typical environment, neutrino mean free path is so large that they do not make up matter. At the core of the supernova, however, the matter density becomes so high that the neutrino mean free path can become much smaller. One can indeed show that the neutrino mean free path can be of order 1 cm when the nuclear matter density becomes of order $10^{15} \ {\rm g}/{\rm cm}^3$ \cite{Yamamoto:2015gzz}. This is much smaller than the typical astrophysical length scale; recall that the typical radius of the core is of order 100 km. This means that the hydrodynamic regime is achieved even for neutrinos, at least at the core of the supernova. The key point that has not been appreciated so far is that neutrino matter there is a {\it chiral quantum liquid} with giant parity violation. The near $\beta$ equilibrium and charge neutrality condition require that the neutrino chemical potential is of order 200 MeV. As the maximal temperature of supernovae (roughly 10 MeV) is much smaller than the neutrino chemical potential, neutrinos are regarded as nearly Fermi degenerate at the core. To the best of our knowledge, this is the only place where condensed matter physics of neutrinos can be relevant in the Universe. Outside the core where the matter density is not sufficiently high, the hydrodynamic description for neutrinos may break down. In such a situation, one needs to use the chiral kinetic theory for neutrinos instead.

As a starting point and for simplicity, below we will consider the pure neutrino matter in the hydrodynamic regime in 3D. This is aimed to understand the qualitative change of the hydrodynamic behavior due to the chirality of neutrinos. Of course, as we argued above, there are not only neutrinos, but also ordinary matter in reality. Moreover, neutrino matter is out of the hydrodynamic regime in the outer region of the supernova. Eventually, one needs to perform numerical simulations of the neutrino radiation chiral hydrodynamics discussed above to describe the whole evolution of supernovae more realistically. Hence, the following discussion on the pure neutrino matter should be considered as a simplified model.

\subsection{Chiral hydrodynamics for neutrino matter}
If one is interested in the dynamics at the macroscopic scale much larger than the mean free path, the chiral kinetic theory reduces to the chiral hydrodynamics. The chiral hydrodynamic equations for pure neutrino matter in the Landau frame are \cite{Son:2009tf}
\begin{equation}
\d_{\mu}T^{\mu \nu} = 0, \qquad \d_{\mu} j^{\mu} = 0,
\end{equation}
where
\begin{align}
\label{T}
T^{\mu \nu} &= (\epsilon + P)u^{\mu} u^{\nu} - P g^{\mu \nu} + \tau^{\mu \nu},
\\
\label{j}
j^{\mu} & = n u^{\mu} + \kappa \omega^{\mu}.
\end{align}
with $\epsilon$ being the energy density, $P$ the pressure, $u^{\mu}\equiv \gamma(1, {\bm v})$ the fluid velocity, $n$ the particle number density, and $\tau^{\mu \nu}$ the dissipative term. As we discussed before, the contribution to the current proportional to the vorticity $\omega^{\mu} \equiv \epsilon^{\mu \nu \alpha \beta} u_{\nu} \d_{\alpha} u_{\beta}$ with the transport coefficient \cite{Son:2009tf},
\begin{equation}
\kappa = -\frac{\mu^2}{8\pi^2} \left(1-\frac{2}{3} \frac{n\mu}{\epsilon+P} \right) 
- \frac{T^2}{24} \left(1- \frac{2 n \mu }{\epsilon+P} \right) \,,
\end{equation}
is the CVE. Note that the expression of $\kappa$ is different from Eq.~(\ref{CVE}) as we take the Landau frame here.

Let us look at the properties of the chiral hydrodynamics. In supernovae, the typical bulk fluid velocity is sufficiently small compared with the speed of light before and after the core bounce. So we can expand the chiral hydrodynamic equations above in terms of $v \equiv |{\bm v}|$ as  \cite{Yamamoto:2015gzz, Yamamoto:2016xtu}
\begin{gather}
\label{hydro1}
(\epsilon + P)(\d_t + {\bm v} \cdot {\bm \nabla}) {\bm v} = - {\bm \nabla} P + \nu {\bm \nabla}^2 {\bm v},
\\
\label{hydro2}
\d_t(n + \kappa {\bm v} \cdot {\bm \omega}) + {\bm \nabla} \cdot {\bm j} = 0.
\end{gather}
Performing the volume integral of Eq.~(\ref{hydro2}), we get the global conservation law:
\begin{equation}
\frac{\rm d}{{\rm d}t} \left(\int {\rm d}^3{\bm x} n + \int {\rm d}^3{\bm x} \kappa {\bm v} \cdot {\bm \omega} \right) = 0\,,
\end{equation}
which stands for the total helicity conservation. The first term is the total neutrino number and the second term is the so-called fluid helicity. The fluid helicity is a parity-odd quantity and expresses the helicity of the fluid. This relation shows that the total neutrino number is not necessarily conserved and can be converted to fluid helicity \cite{Yamamoto:2015gzz, Yamamoto:2016xtu}. 

One can also show that these hydrodynamic equations have the following scaling symmetry in the inertial range (where the effects of dissipation can be ignored):
\begin{gather}
\label{symmetry_n}
{\bm x} \rightarrow l{\bm x}, \quad t \rightarrow l^{1-h}t, \quad {\bm v} \rightarrow l^{h}{\bm v}, \quad \mu \rightarrow l^p \mu,
\end{gather}
both for $\mu \gg T$ and $\mu \ll T$. In the presence of the CVE, the scaling exponents $h$ and $p$ are uniquely fixed as \cite{Yamamoto:2016xtu}
\begin{equation}
\label{hp}
h=0, \qquad p=-1.
\end{equation}

\subsection{Chiral turbulence}
Let us discuss the consequences of this unique scaling symmetry. Consider the spectrum of the fluid kinetic energy in momentum space defined by
\begin{equation}
\label{E_v}
{\cal E}_v(k, t) = \frac{2\pi k^2}{(2\pi)^3} \int {\rm d}^3 {\bm y} \ e^{i {\bm k} \cdot {\bm y}}
\langle {\bm v}({\bm x}, t) \cdot {\bm v}({\bm x} + {\bm y}, t) \rangle \,.
\end{equation}
Then the scaling symmetry (\ref{symmetry_n}) leads to the relation ${\cal E}_v(l^{-1}k, l^{1-h}t) = l^{1+2h} {\cal E}_v(k, t)$. One can define the scaling function $\psi_v(k,t) \equiv k^{1+2h}{\cal E}_v(k,t)$, such that $\psi_v(l^{-1}k,l^{1-h}t)=\psi_v(k,t)$. The correlation length of the fluid energy can also be defined as
\begin{equation}
\label{xi_v}
\xi_v (t) = 2\pi \frac{\int_0^{\infty} {\rm d}k \ k^{-1}{\cal E}_v(k,t)}{\int_0^{\infty} {\rm d}k \ {\cal E}_v(k,t)}\,.
\end{equation}

We assume that the turbulent behavior approaches the scaling solution above independently of the initial conditions after sufficiently long time. Inserting the scaling exponent $h$ in Eq.~(\ref{hp}), the asymptotic behavior of $\xi_v(t)$ is then determined as \cite{Yamamoto:2016xtu}
\begin{equation}
\label{xi_v_scaling}
\xi_v(t) =\xi_v(t_s) \left(\frac{t}{t_s} \right) \,,
\end{equation}
where 
\begin{equation}
\xi_v(t_s) = 2\pi t_s \frac{\int_0^{\infty} {\rm d}x \ x^{-2} \psi_v(x)}{\int_0^{\infty} {\rm d}x \ x^{-1} \psi_v(x)}\,,
\end{equation}
with $t_s$ being some parameter. Therefore, one finds that the correlation length grows as a function of time. This is the {\it inverse energy cascade}: the fluid kinetic energy is transferred from a small scale to a larger scale. 

It is known that usual parity-invariant matter exhibits the {\it direct energy cascade} in 3D, where the turbulent energy flow goes from a large scale to a smaller scale. The direct energy cascade has actually been observed in the numerical simulations of the conventional neutrino transport theory in 3D {\it without} taking into account the chirality of neutrinos \cite{Hanke:2011jf}. Our discussion above shows that the effects of the chirality of neutrinos can reverse the direction of the turbulent energy cascade. This feature is potentially important in the origin of the supernovae explosion: because the inverse energy cascade can generate a large-scale coherent fluid motion, the chiral effects of neutrinos can work favorably for the supernova explosion. It should be remarked once again that we have here considered a simplified model, and whether this scenario is the case in more realistic situations should be checked with the neutrino radiation chiral hydrodynamics \cite{radiation}. 

In this context, it is interesting to note that even usual 3D Navier-Stokes equations show the inverse turbulent energy cascade when only the modes with a given sign of helicity are retained in the helical decomposition of fluids \cite{inverse}. This suggests that the effects of parity violation (and consequently, the presence of fluid helicity) play key roles in the hydrodynamic evolution of a broader class of systems.

\section{Conclusion}
We argued the importance of chiral transport of neutrinos in supernovae. The chirality of neutrinos modifies the macroscopic hydrodynamic evolution of supernovae and provides a tendency toward the inverse energy cascade even in 3D. Moreover, the chiral effects, coupled with electromagnetic fields, may naturally generate strong magnetic field with magnetic helicity through the chiral plasma instability \cite{Akamatsu:2013pjd}, providing a possible mechanism of magnetars \cite{Yamamoto:2015gzz}. It has also been suggested that the chiral effects may potentially explain the neutron star kicks \cite{Kaminski:2014jda}. It is important to test these possible mechanisms by the numerical simulations of the neutrino radiation {\it chiral} hydrodynamics \cite{radiation} in the future.

\section*{Acknowledgement}
This work was supported by JSPS KAKENHI Grants No.~16K17703 and MEXT-Supported Program for the Strategic Research Foundation at Private Universities, ``Topological Science" (Grant No.~S1511006).

\end{document}